# Enhanced flexural wave sensing by adaptive gradient-index metamaterials

Y.Y. Chen[1], R. Zhu[1], M.V. Barnhart[1] & G.L. Huang[1]

Increasing sensitivity and signal to noise ratios of conventional wave sensors is an interesting topic in structural health monitoring, medical imaging, aerospace and nuclear instrumentation. Here, we report the concept of a gradient piezoelectric self-sensing system by integrating shunting circuitry into conventional sensors. By tuning circuit elements properly, both the quality and quantity of the flexural wave measurement data can be significantly increased for new adaptive sensing applications. Through analytical, numerical and experimental studies, we demonstrate that the metamaterial-based sensing system (MBSS) with gradient bending stiffness can be designed by connecting gradient negative capacitance circuits to an array of piezoelectric patches (sensors). We further demonstrate that the proposed system can achieve more than two orders of magnitude amplification of flexural wave signals to overcome the detection limit. This research encompasses fundamental advancements in the MBSS with improved performance and functionalities, and will yield significant advances for a range of applications.

[1]Department of Mechanical and Aerospace Engineering, University of Missouri, Columbia, MO, 65211, USA. Correspondence and requests for materials should be addressed to G.L.H. (email: huangg@missouri.edu).

Elastic (mechanical) wave sensors play a crucial role in the detection of elastic wave signals for measuring and imaging structural dynamic behaviors in order to ensure reliability and detect possible damage[1-4]. Applications include structural health monitoring[5], medical imaging[6], aerospace[7] and nuclear instrumentation[8]. Piezoelectric sensors are one of the most widely employed devices in current elastic wave sensing systems due to their universal functionality, compactness, high bandwidth capabilities and easy integration with existing host structures. The performance is determined by the dynamic stiffness (or the impedance of the sensor system) and generalized electro-mechanical coupling factor. However, the detection limit and sensitivity issues still hinder the performance of these sensors especially in large structures where weak signals or low signal to noise ratios (SNRs) exist. Therefore, it is still a necessity to develop a sensing system with the proper dynamic stiffness that can overcome the detection limit, improve the sensitivity and increase the SNRs of conventional piezoelectric-based sensors.

Much like their optical counterparts, acoustic/elastic metamaterials, which are artificially engineered materials, have shown promising potential for manipulating acoustic/elastic waves at the subwavelength scale. Innovative methods for acoustic/elastic wave collimation, focusing, cloaking, super-resolution imaging, negative refraction and wave steering have been explored by tailoring gradient effective material properties including positive/negative mass density and modulus either independently or simultaneously[9-17]. An acoustic rainbow trapping metamaterial which consisted of an array of grooves with graded depths perforated on a rigid plate to control the spatial distribution of the acoustic energy was recently demonstrated numerically and experimentally[18]. It was found that graded acoustic meta-structures can control not only the propagation directions of the waves but also the spatial distributions of the energy. The gradient meta-structure was further interpreted and modelled as an acoustic metamaterial with strong anisotropy of effective mass density and the intensity of acoustic field was effectively enhanced near the propagation-stop position due to the slow group velocity. Recently, by effectively trapping broadband acoustic waves along the waveguide, enhanced acoustic sensing was achieved by detecting the significantly amplified pressure field in such a high-refractive-index medium to overcome the detection limit in conventional acoustic wave sensors[19]. However, in stark contrast to the tremendous progress in acoustic metamaterials, decelerating, trapping and spectrum-splitting of elastic waves by using elastic metamaterials has not yet been realized due to the fact that there is still a lack of a feasible and easily tunable meta-structure with strong gradient index (GRIN) and wave dispersion. These abilities are critical in many applications ranging from elastic wave spatial focusing to frequency selections in elastic wave detection and imaging.

The piezoelectric shunting technique is one well known method that receives a considerable amount of attention. Adaptive structural systems with integrated piezoelectric shunt circuits have been proven as a promising method to achieve mechanical vibration/wave attenuation and control by electrically tuning the dynamic stiffness, mass density and impedance[20-32]. This technique is convenient because the transducer is both a sensor and an actuator but with a reduced weight. Lightweight piezo-patches periodically shunted by passive inductance-resistance (LR) electrical circuits can produce locally resonant band gaps and moreover, piezo-patches with semi-active negative capacitances (NCs) allow the tunable band gaps or wave-guiding over desired broadband frequency ranges through the use of integrated active components in circuits[33-36]. More recently, a design for an adaptive elastic metamaterial was proposed and demonstrated numerically as well as experimentally by integrating the shunted NC piezoelectric patches with uniform circuits into the metastructure[37-39]. The tunable bandgap behavior for wave frequency selections was demonstrated experimentally for the first time in a deep subwavelength scale without physical microstructural modifications. The underlying principle of the piezoelectric shunting circuitry concept is that the additional electric degree-of-freedom and associated dynamic behavior can be tailored to enhance the systems performance.

In this paper, we combine the concepts of the acoustic rainbow trapping metamaterial and adaptive systems with piezoelectric circuitry to demonstrate sensitivity enhancing concepts of elastic waves by using an array of piezoelectric elements shunted with non-uniform gradient circuits. Furthermore, we design and fabricate a metamaterial-based sensing system that is capable of amplifying broadband flexural waves by electronically tuning the effective bending stiffness gradient. Without loss of generality, we demonstrate a one-dimensional (1D) GRIN metamaterial beam (Fig. 1), where high GRINs can be tailored with different circuits. When a flexural wave propagates into an elastic medium with a highly increased GRIN, the wavelength is shortened due to the increased wavenumber $k$, meaning the elastic strain energy is concentrated and amplified spatially. Therefore, enhanced elastic wave sensing can be achieved by detecting the significantly amplified strain field in the high-refractive-index medium. In addition, different from passive approaches, the electrical signal of the piezoelectric sensors with adaptive circuits is also amplified such that the mechanical sensing performance will be electrically enhanced. In conjunction with the adaptive piezoelectric circuitry, this study seeks to demonstrate a new hybrid technique for enhancing mechanical wave sensing which is robust when response signals are distorted due to realistic noise and damping factors which otherwise degrade the sensitivity of conventional piezoelectric sensors. As proof of concept, we demonstrate flexural wave propagation and compression in one type of high stiffness-gradient-index metamaterial constructed with an elastic beam with an array of the piezoelectric sensors with gradient shunting circuits. To understand the wave compression and amplification phenomenon caused by the electro-mechanical coupling, a multi-physics analytical model is developed to capture the spatial and spectral wave distributions in the metamaterial. In our experiments on the prototype metamaterial device, flexural wave amplification of more than two orders of magnitude is demonstrated, which is further validated by using numerical simulations. To illustrate the enhanced flexural wave sensing in the metamaterial device, a transient flexural wave signal overwhelmed by noise is successfully recovered to overcome the detection limit of conventional piezoelectric sensors.

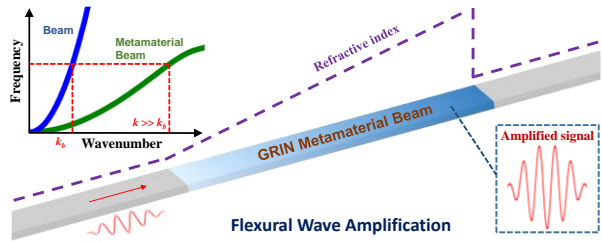

**Figure 1 | Schematic of the GRIN metamaterial-enhanced flexural wave sensing.** Bottom: the strain field of the flexural wave is amplified inside the high-refractive-index metamaterial beam. Top left: schematic of metamaterial dispersion curves for flexural waves with a high wavenumber (refractive index) state.

## Results

**Theoretical modeling of the adaptive GRIN metamaterial beam for flexural wave control.** The introduction of smart materials into metamaterial building blocks has the potential to provide a novel method for controlling the effective material properties. In this study, shunted piezoelectric elements are used as variable stiffness elements in order to provide such control. This approach is based on a combination of the inherit properties present in the metamaterial concept with the addition of shunted piezoelectric materials which can be used to alter material properties in real-time. Recently, Chen et al.[37,39] implemented NC piezoelectric shunting technique into a metamaterial to actively tune the local resonant (LR) frequency by modifying the stiffness of the resonant microstructure and therefore, controlling the band gap of the active elastic metamaterial (EMM). One advantage of a NC shunted circuit is the ability to continuously modify the resonant properties of the metamaterial over a broad frequency range.

As illustrated in Fig. 2a, the adaptive GRIN metamaterial waveguide is constructed by bonding a periodic array of piezoelectric (PZT-5A) patches to an aluminum beam, and the total length of the waveguide is denoted as $L$. The patches are individually shunted with an array of gradient negative capacitors to ensure a linear variation in the refractive index. Different GRINs can also be easily tuned, but for simplicity and without loss of generality, a linear variation is selected in this study. Figures 2b and 2c show the $n$-th unit cell of the adaptive metamaterial beam and the schematic of its shunted NC circuit, respectively. The length and thickness of the beam and piezoelectric patch are represented by $L_b$, $h$, $L_p$ and $h_p$, respectively. All of the PZT patches are polarized in the positive $z$-direction and electrodes are attached to the upper and lower surfaces. The shunted NC is denoted by $-c_N$, which is produced by an operational amplifier, two resistors ($R_0$ and $R_2$), a capacitor ($C_0$), and a potentiometer ($R_1$). By assuming the operational amplifier to be ideal, and the impedance of $R_0$ is large enough compared with $C_0$, the equivalent NC of the circuit can be written as

$$-c_N = -\frac{R_1}{R_2}c_0. \quad (1)$$

Therefore the NC can be easily tuned by adjusting $R_1$ in a stable circuit operation region. One distinct advantage of this design is that the inherent voltage amplification induced by the NC circuit can also enhance sensor signals which are measured from the shunting circuits. Therefore, the total sensor enhancement is caused by the collective effects from both the wave compression due to the GRIN metamaterial and internal circuit amplification. In order to quantitatively characterize the hybrid wave amplification and propagation behavior, a 1D multi-physics analytical model is developed based on the plane stress assumption (see supplementary material for details).

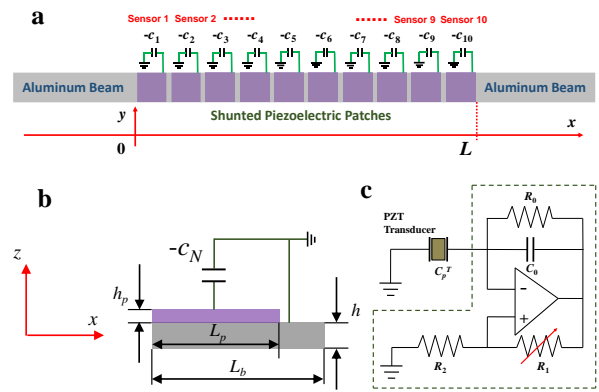

**Figure 2 | Schematic of the adaptive GRIN metamaterial beam design.** (**a**) An aluminum beam is bonded by a periodic array of piezoelectric (PZT-5A) patches, which are shunted by a gradient array of NC circuits to realize an adaptive GRIN metamaterial waveguide. Each piezoelectric patch with its shunting circuit also serves as an individual sensor. The sequence of sensors is numbered from the left hand side to the right hand side of the waveguide. (**b**) Schematic of the unit cell of the adaptive metamaterial. For simplicity, in the piezoelectric coupled analytical model, the plane stress assumption is considered for the adaptive metamaterial. The propagation of flexural waves is assumed to be confined along the $x$ direction. (**c**) Schematic of the NC shunting circuit, which is produced by an operational amplifier, two resistors, a capacitor, and a potentiometer. By adjusting the resistance of the potentiometer, the equivalent NC of the circuit can be altered.

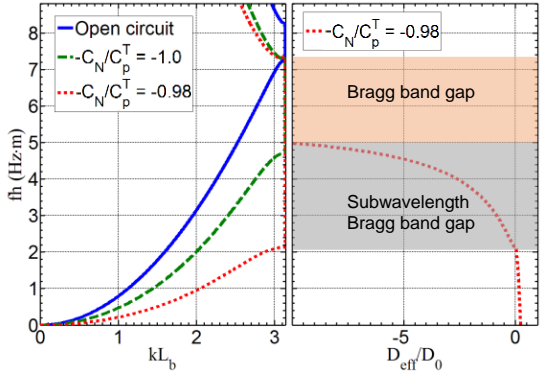

**Figure 3 | Flexural wave dispersion relations and the related effective bending stiffness of the adaptive metamaterial.** Left: Flexural wave dispersion relations of the adaptive metamaterial with different uniformly distributed shunting circuits. Right: effective bending stiffness of the adaptive metamaterial when $-c_N/c_p^T = -0.98$. Shaded areas denote the subwavelength and regular Bragg band gaps, respectively.

To demonstrate NC effects, flexural wave dispersion relations and the related effective bending stiffness of the adaptive metamaterial are plotted in Fig. 3 for the PZT patches with open circuits and uniform NC circuits with $–c_N/c_p^T = -1.0$ and -0.98, respectively, where $c_p^T$ is the capacitance of the PZT patch at constant stress (see supplementary material for details). The material and geometric parameters are listed in Table. 1. As shown in Fig. 3, the frequency range of the band gap will become much lower (subwavelength range) and broader when the NC circuit is used, compared with the case with open circuits. Specifically, when the PZT patches are connected with NC shunting circuits, i.e. $–c_N/c_p^T = -1.0$ or -0.98, the frequency ranges of the band gap are changed to the non-dimensional regions of 4.75 Hz·m – 7.35 Hz·m or 2.15 Hz·m – 7.35 Hz·m, respectively, from the one of 7.35 Hz·m – 8.25 Hz·m for the open circuit. This shifting phenomenon can be explained by the effective bending stiffness control due to the shunting circuits, which is calculated with the reaction bending moment of the unit cell from the dispersion analysis (see supplementary material for details). As illustrated in the right hand side window of Fig. 3, for the metamaterial with uniform NC circuits being $–c_N/c_p^T = -0.98$, the total broad band gap is composed of the subwavelength Bragg scattering, which is interpreted by the negative bending stiffness, and conventional Bragg band gap.

The negative bending stiffness of the metamaterial is mainly caused by the significantly mechanical impedance mismatch between the negative effective modulus of the PZT patches and the host beam. The lower edge of the subwavelength Bragg band gap frequency range coincides with the frequency where the effective bending stiffness switches from a positive to a negative value. For the case of $–c_N/c_p^T = -1.0$, the reduced effective modulus of the shunted PZT patches is not very significant and the conventional Bragg scattering is therefore the dominant wave attenuation mechanism. This strong tunability of the effective modulus of the PZT patches by using NCs will be adopted for the design of the adaptive GRIN metamaterial in order to manipulate wave propagation, amplification and band gaps at the subwavelength scale.

**Table 1 | Geometric and material properties of the adaptive metamaterial.**

| Geometric properties (in mm) | | Material properties (Al) | |
|---|---|---|---|
| $L_b$ | 7.4 | Mass density | 2700.0 kg/m³ |
| $h$ | 0.4 | Young's modulus | 69.0 GPa |
| $L_p$ | 7.0 | Poisson's ratio | 0.33 |
| $h_p$ | 0.25 | | |
| **Material properties (PZT-5A)** | | | |
| Mass density | | | 7750.0 kg/m³ |
| Elastic compliance ($s_{11}^E$) | | | 16.4×10⁻¹² m²/N |
| Dielectric constant ($\varepsilon_{33}^T$) | | | $1700.0\varepsilon_0$ |
| Piezoelectric constant ($d_{31}$) | | | 171.0×10⁻¹² pC/N |

To further demonstrate the flexural wave amplification, a linear GRIN of the metamaterial along the flexural waveguide is first realized analytically. Based on the analytical model of dispersion relations, different GRINs can be tuned by altering the circuit parameter $R_1$ in the adaptive metamaterial. As an example, shown in Fig. 4a, the maximum refractive index is as high as $n_{max} = 5$ and the minimum refractive index is as small as $n_{min} = 1.4$ by properly selecting gradient NCs shunted to the PZT patches, properties of which are hard to find in naturally occurring materials. Wave displacement distribution fields in the adaptive GRIN metamaterial under harmonic loading are illustrated in Fig. 4b over different frequencies. As shown in the figure, it is evident that the flexural wavelength gradually decreases in the adaptive GRIN metamaterial waveguide at all three frequencies, which renders a spatial compression of the flexural wave. Such wave compression due to the GRIN is one of the major mechanisms necessary to enable strain amplification within the waveguide. Furthermore, it should be mentioned that the GRIN metamaterial with the linear GRIN will also produce a gradient shift of band gaps to the lower subwavelength frequency region, which can lead to a gradient cutoff frequency along the waveguide (solid curve in Fig. 4a). Therefore, lower frequency waves can propagate over greater distances along the waveguide (i.e. $fh = 0.2$ Hz·m), however, higher frequency waves will be reflected back in the front portion of the designed waveguide (i.e. $fh = 2.0$ Hz·m) as shown in Fig. 4b. It can also be seen that the cutoff frequencies are located at different positions on the metamaterial meaning the metamaterial is capable of frequency multiplexing. The maximum wave amplification is expected at the cutoff frequencies while the flexural wave is prohibited above these frequencies. To quantitatively investigate the amplification effect due to the GRIN, the strain gain at various positions, which is defined by the ratio between the normal strain amplitude on the GRIN metamaterial beam and the normal strain amplitude on the host beam along the waveguide, is obtained for different frequencies in Fig. 4c. As shown in the figure, the strain gain gradually increases along the waveguide until the flexural wave is reflected back due to the cutoff frequencies at different locations. The constant strain gain regions are the locations occupied by the PZT patches where the constant electric charge and voltage on the upper electrode of each PZT patch is induced by the NC shunting circuit. Specifically, for the normalized frequency $fh = 0.2$ Hz·m, the wave cutoff position is around $x/L = 0.8$ and the highest strain gain can be around 18. However, the wave cutoff position is shifted to $x/L = 0.5$ and the highest strain gain is around 7 when the normalized frequency is $fh = 0.6$ Hz·m. To obtain a greater mechanical stain gain for higher frequency waves, the size of PZT patches and the lattice constant of the metamaterial should be reduced. However, for the shunted PZT patches serving as wave sensors, the final wave amplification is revealed in the measured electric signals or amplified voltages. This amplified voltage at the PZT patch can be represented by the voltage gain, which is defined by the ratio between the potential amplitude with the NC circuit and the potential amplitude with open circuit on the upper electrode of the PZT patch as (see supplementary material for details)

$$G_V = \frac{G_s}{1 + \dfrac{1}{i\omega C_p^T \left(1 - k_{31}^2\right) Z_{sh}}}. \quad (2)$$

where $G_s$ and $k_{31}$ denote the averaged strain gain of the PZT patch between the shunting circuit and open circuit conditions and its electromechanical coupling coefficient, respectively, and $Z_{sh}$ represents the impedance of the shunting circuit. According to equation (5), the voltage gain is the final parameter needed to capture the hybrid wave amplification effects from both the strain gain and the shunting circuit enhancement. It should be clarified that the signal amplification through NC circuit is different from conventional charge or voltage amplification conditioning circuits. The NC circuit can enhance the strain in the PZT sensors through voltage control and the enhanced strain will increase the voltage in the PZT patch accordingly. This indicates that an extremely large voltage gain could be created by the NC circuit and this feedback effect will also increase the sensitivity of the conventional sensors. However, conventional amplification circuits can only amplify electric signals but lack of the feedback control such that there will not be any improvement in the sensor sensitivity. Figure 4d shows the calculated voltage gains for different PZT patches when harmonic waves propagate along the metamaterial waveguide. First, the voltage gains are in qualitative agreement with the strain gains along the waveguide for different frequencies. However, the amplification magnitude for the voltage is significantly higher than that of the strain where more than two orders of magnitude amplification is observed. For example, the voltage amplification magnitude is as high as 750 at the location $x/L = 0.7$ for the normalized frequency $fh =$ 0.2 Hz·m. It can be concluded that the NC shunting circuit will play another important role for sensor enhancement and can provide surprisingly high enhancement for the sensor signals when compared with the GRIN effect, which has never been studied or demonstrated before. It is also observed that the amplification magnitude and location are also dependent on the propagating wave frequencies, and optimization analysis is needed in order to consider this trade-off effect by properly designing the GRIN of the metamaterial for different wave frequencies. To further quantitatively illustrate the hybrid sensing system of the proposed metamaterial waveguide, the amplification behavior of the strain and voltage in the frequency domain at different sensor locations is investigated. For clear demonstration, the mechanical strain and total voltage gains at the 4$^{th}$, 6$^{th}$, 8$^{th}$ and 10$^{th}$ PZT patches along the waveguide are plotted in Figs. 4e and 4f, respectively. It is evident that the total voltage gain of up to 3500 can be achieved compared with the highest mechanical strain gain being around 45 at the 10$^{th}$ sensor (the rear location of the waveguide). This is expected because the largest wave compression is obtained when the wave propagates through the metamaterial with the highest refractive index. However, for subwavelength frequencies in the current GRIN metamaterial waveguide design, the sensing enhancement frequency range is very narrow because of the frequency cutoff due to the GRIN. It is also noted that the 4$^{th}$ or 6$^{th}$ sensors (the front location of the waveguide) can achieve enhanced sensing in the high frequency and broadband regions although the gain is reduced. Therefore, this is also a trade-off between the sensing gain and bandwidth; as the gain increases, the bandwidth decreases. In addition, it is worth noting that other geometric parameters such as a material mismatch between the PZT patches and host medium and width of the metamaterial waveguide can also influence the flexural wave gain. It is expected that optimization of the metamaterial's geometry and gradient parameters can lead to even better performance, which will be investigated in the near future.

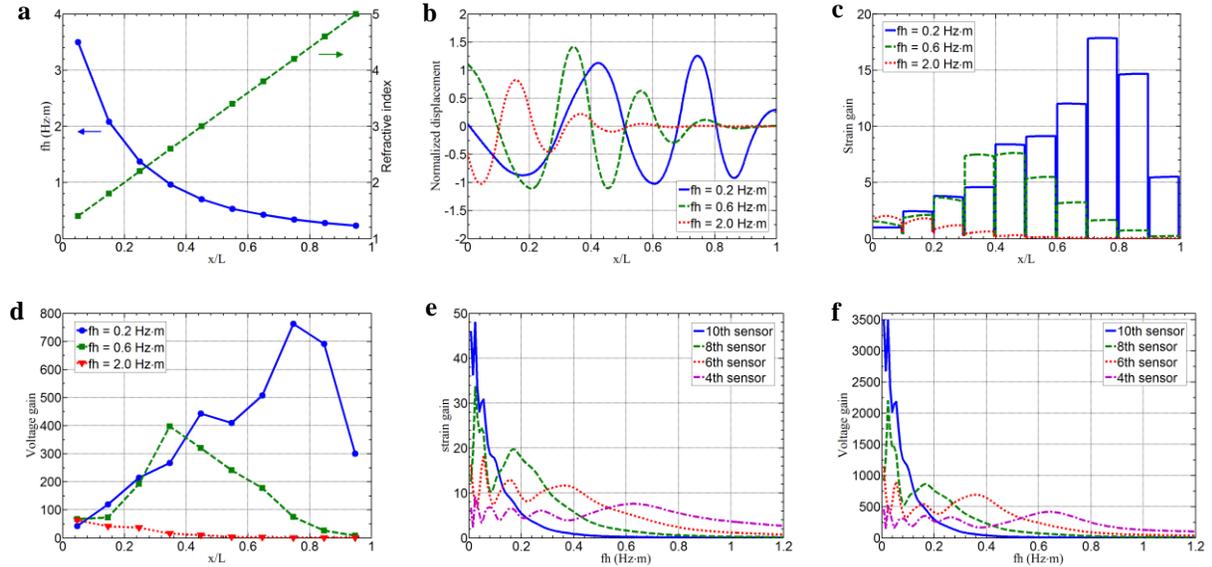

**Figure 4 | Flexural wave compression and amplification effects of the adaptive metamaterial based on analytical model.** (**a**) The flexural wave refractive index and corresponding cutoff frequencies of each unit cell along the adaptive GRIN metamaterial waveguide. (**b**) The compressed transverse displacement field along the adaptive waveguide at different frequencies. (**c**) The strain gain on the waveguide along the wave propagation direction for different frequencies. (**d**) The voltage gain in the sensors along the wave propagation direction for different frequencies. (**e**) The frequency spectra of the strain gain at different sensor locations on the metamaterial. (**f**) The frequency spectra of the voltage gain for different sensors on the metamaterial.

**Numerical simulations of GRIN metamaterial devices.** For the real GRIN metamaterial device, when the flexural wave propagates through a finite waveguide beam, the boundary effects on the electromechanical coupling can be significant, which is not considered in the analytical model. Here, we perform multiphysics numerical simulations that consider piezoelectric and shunting circuit effects by using the commercial software COMSOL Multiphysics. In order to find the proper NCs shunted into each PZT patch to create a desired GRIN metamaterial waveguide, wave dispersion analysis based on the 3D linear piezoelectric theory and Bloch theorem are conducted numerically[39], from which the complex wavenumber of flexural waves can be calculated with a given frequency. Through some simple iterative procedures, the relationship between the refractive index and NC can be found at specific frequencies. After this, the linearly GRIN metamaterial waveguide can be designed. The wave compression and signal amplification performance are examined in both frequency and time domains by using the piezoelectric modulus and considering the shunting circuit effects. The time domain analyses are performed with the real geometric and material properties used during the experiments. The numerical results are then used to guide the circuit tuning in the experiment based on the relation between voltage signal amplitudes and shunted NC values.

**Experiments on flexural wave amplifications.** Based on the flexural wave amplification demonstrated analytically and numerically in the GRIN metamaterial, a metamaterial-enhanced flexural wave sensing system is developed experimentally. The adaptive GRIN metamaterial is fabricated by surface-bonding seven equally-spaced PZT patches shunted with gradient NC circuits on the one side of the host beam (see Fig. 5a). To reduce the wave disturbance in the waveguide, the width of the PZT patches and the host beam are made to be equal (being 7mm). In the NC circuit, a 1 nF reference capacitor $C_0$, which is smaller than the capacitance of the PZT patch at constant stress (~3 nF) is selected to induce an additional amplification gain (~×3) in the final measured voltage signals. However, making $C_0$ too small is not recommended, as it can destabilize the real circuit. The resistor $R_0$ used in the NC circuit needs to be carefully chosen based on the operating frequencies. An extremely large $R_0$ value can make the circuit unstable at relatively large NC values, which usually cannot satisfy the large GRIN applications. At the same time, if the value of $R_0$ is too small it will not be acceptable for synthetizing the NC circuit. In our experiment, a 1 MΩ resistor is selected for $R_0$. The gradient NC circuits are individually connected to the seven PZT patches and are powered by a DC power supply. By adjusting the potentiometer $R_1$ in the NC circuit, the value of NC is gradually varied, and the gradient effective modulus of the metamaterial can be tuned electronically. Based on the amplified voltage signal gain of each sensor (other sensors have open circuits) in the GRIN metamaterial waveguide from the time domain numerical studies, each of the individual NC circuits are carefully tuned (while the other PZT patches have open circuits) to ensure a desired GRIN. In order to implement this in the real NC circuits, the stability issue must be considered during the experiments. Therefore, the maximum refractive index in the flexural waveguide is selected to be 2.1 at the 7th sensor to avoid approaching the stability boundary of the system and ensure stable experimental operation. Although this index is not as high as that used in the analytical method, it will be sufficient for wave signal amplification that will be discussed subsequently. In the experiment, we excite the flexural waveguide using an additional PZT patch actuator attached approximately 35 mm away from the metamaterial beam's origin. The piezoelectric element was excited using a tone burst signal with a duration of five cycles. This kind of excitation allowed us to bundle energy in the frequency range of interest while having shorter measurement sequences than we would have needed using a chirp signal. The amplified voltage signals are directly measured by connecting a digital oscilloscope to the output terminals of the operational amplifiers in the NC circuits. The out of plane velocity field on the surface of the sample is measured by using a scanning laser vibrometer.

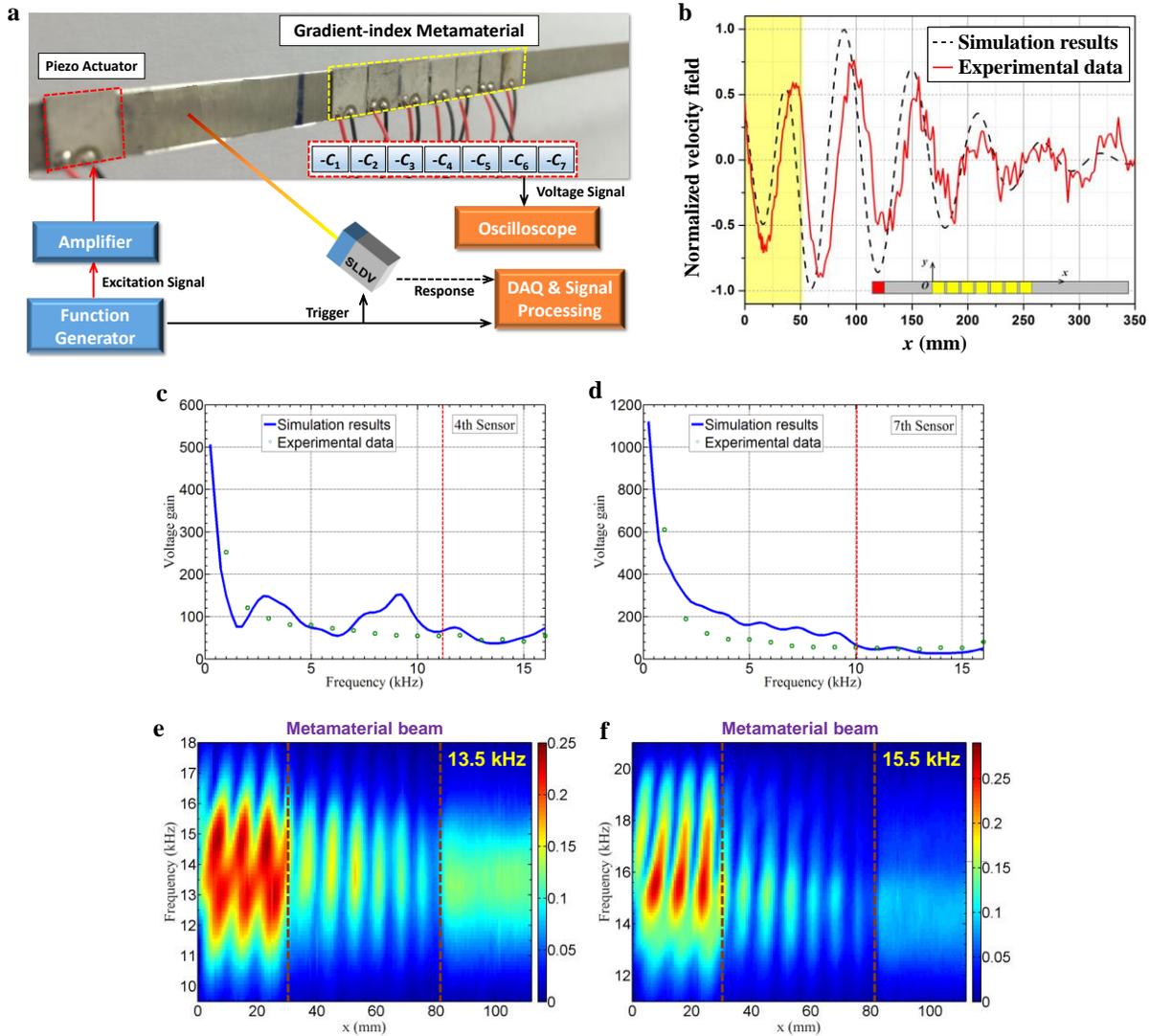

**Figure 5 | Experimental measurements of the adaptive metamaterial response.** (**a**) Schematic of the experimental setup of the adaptive metamaterial enhanced sensing system. (**b**) Experimental and numerical characterizations of the transverse displacement fields along the waveguide at $t = 2.98$ ms. The solid and dashed curves represent the experimental data and numerical results, respectively. (**c**) Voltage gain spectra measured at the 4th sensor. (**d**) Voltage gain spectra measured at the 7th sensor. In **c** and **d**, the circles denote the directly measured data, and the solid curves represent results from numerical simulations. (**e,f**) Wave velocity amplitude as a function of frequency and location along the waveguide with central frequency of the incident signal being 13.5 and 15.5 kHz, respectively.

The normalized transient displacement fields of the flexural waves on the central line of beam's bottom surface at the same time ($t = 2.98$ ms) are illustrated in Fig. 5b for the experimental testing and numerical simulation, where the incident signal's central frequency is 1 kHz. The shaded region denotes the location of the GRIN metamaterial. Wave compression before and after the GRIN metamaterial can be clearly observed along the waveguide with a reduced wavelength in this region displayed in both the numerical and experimental results, which is consistent with the theoretical prediction (Fig. 4b). As shown in the figure, the wave compression from experimental measurements is weaker than that from the numerical simulations, and the difference can be attributed to the fact that the inherent resistance from the resistor used in parallel with the reference capacitor, wires, connection boards and operational amplifiers in the NC circuits are not taken into account in simulations. Flexural wave amplification in terms of the voltage gain are demonstrated in the frequency spectrum at the 4th and 7th sensors as shown in Figs. 5c and 5d, respectively. We carry out a Fast Fourier transform of the time-domain data for the voltage signals along the length of the sample to obtain the frequency response. The numerical investigations are carried out in the frequency domain, and are in good agreement with the time domain analysis performed conceptually in the same way with the experiments. As shown in the figures, the frequency spectra of the voltage gain in our experiments agrees reasonably well with those predicted using numerical simulations, and voltage amplifications of more than two orders of magnitude are achieved for broadband frequencies. The differences between the numerical and experimental results are due to the non-dissipative mechanical and electrical properties assumed in the simulations, therefore, the reflected flexural wave within the waveguide will overlap with the incident wave and make the voltage gain fluctuate with frequency. However, during the experiments the inherent mechanical and electrical dissipation cause the reflected flexural wave to become weaker and make the voltage gain fluctuation more steady in relation to the frequency. The cutoff frequencies of the 4th and 7th sensors are numerically determined to be 10.04 and 11.18 kHz (denoted as the dashed lines in figures), respectively. Unfortunately, the cutoff behavior on the voltage gain cannot be observed clearly in the figures. This is because the GRIN value is low in the experiments, which is limited by the current NC circuits, and the wave attenuation ability is therefore not significant due to the conventional Bragg scattering effects. Therefore, the leaking waves propagating along the waveguide still possess the amplification property, which is different from the resonant based mechanism in acoustic rainbow trapping. For this enhanced waveguide sensing system, it is also observed that the voltage gain at the 7th sensor is much higher than that in the 4th sensor at frequencies below 3 kHz which is expected because the higher refractive index is located in the rear portion of the GRIN metamaterial. However, for higher frequency waves (above 3 kHz), the voltage gains at the two sensors are still relatively high at around 100 and very close because the voltage gain is mainly caused by the internal circuit amplification. This result is very encouraging for further enhanced sensing systems using only discrete PZT sensors and the amplification is not required to be significantly high.

The frequency cutoff behavior of the waveguide are shown in Figs. 5e and 5f through the frequency spectra of the out of plane velocity as a function of position with central frequencies of incident signals being 13.5 and 15.5 kHz, respectively. The Fast Fourier transform of the time-domain data for the out of plane velocity along the length of the sample is carried out to obtain the frequency response. As shown in both figures, the velocity amplitudes in the waveguide and downstream (within the metamaterial) region are lower than that in the upstream (before the metamaterial) region, which illustrates the wave reflection within the waveguide at these two frequencies. As the wave packet traverses the GRIN metamaterial, the change in wave velocity amplitude as a function of frequency becomes visible, whereas in the 15.5 kHz region where most of the energy is concentrated (as shown by the high amplitudes upstream of the metamaterial), the amplitude is significantly reduced, leading to the very low velocity amplitudes recorded downstream of the metamaterial region. Comparing these two figures, we can also find that the lower frequency wave can propagate over a longer distance into the waveguide and then reflect back, which is consistent with the theoretical prediction.

**Breaking the conventional flexural wave detection limit.** Flexural waves are widely used as mechanical signals in structural health monitoring, energy harvesting and mechanical sensors. In these systems, the detection limit is determined by the noise floor of the mechanical sensors. Here we will demonstrate the feasibility of recovering weak signals overwhelmed by noise using the proposed adaptive metamaterial waveguide system in Fig. 3a. In the hybrid system, both the strain field and measured voltage amplitude are amplified in the wave propagation medium as well as within detection sensors.

To demonstrate this concept, a very weak five peak tone burst signal with a central frequency of 1 kHz is applied to the PZT transducer without using the signal amplifier to generate a propagating flexural wave. The magnitude of the applied voltage is adjusted to control the signal-to-noise ratio (SNR) for the flexural wave signal in the host beam with the conventional PZT sensor. Figures 6a and 6b show the directly measured flexural wave signal and its corresponding frequency amplitudes from the 7th sensor, respectively, with open circuits attached to all of the PZT patches. The flexural wave is completely overwhelmed by noise which means that the signal is smaller than the detection limit of the conventional PZT sensor. The chosen frequency of the flexural wave is 1 kHz so that the wave signal can be amplified within the proposed GRIN metamaterial. As shown in Figs. 6c and 6d for the time and frequency domain signals, respectively, the input tone burst signal is successfully recovered by measuring the signal within the GRIN metamaterial at the 7th sensor and the SNR achieved in the metamaterial is enhanced as high as 100 times. These results demonstrate that the metamaterial-enhanced sensing waveguide system can overcome the detection limit of a conventional elastic sensor.

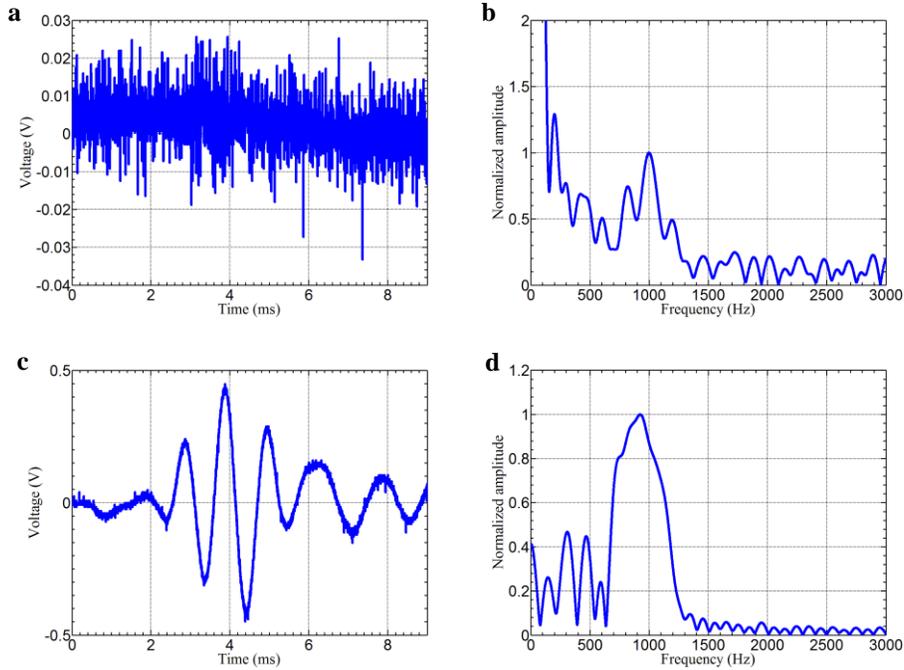

**Figure 6 | Experimental demonstrations of the enhanced sensing capabilities.** (**a**,**b**) Time-domain and frequency-domain flexural wave signal measured from the 7th sensor, when shunting circuits are open. (**c**,**d**) Time-domain and frequency-domain flexural wave signal measured from the NC shunting circuit of the 7th sensor.

**Discussion**

The enhanced flexural wave sensing system with adaptive GRIN metamaterials embraces the wave compression mechanism due to the GRIN and electric signal amplifications within the circuits through the piezoelectric shunting methodology. In our proof-of-concept experiments, amplification of more than two orders of magnitude of flexural wave voltage signals is achieved for extremely broadband frequencies with a maximum refractive index of only 2.1. We anticipate that if the refractive index can be further increased from the current value used in the experiments, unprecedented wave control, compression and amplification of the voltage gain can be accomplished. However, to achieve the high GRIN values, there are several obstacles due to system instability issues related to the NC circuit and limitations of the output voltage level of the operational amplifier. Some geometric and circuit parameter design optimizations are therefore still necessary. Furthermore, by replacing the analog NC shunting circuit with a digital synthetic impedance circuit, the instability issues within the circuit could be theoretically eliminated. Another advantage of the digital implementations is the computation power, from which the signals acquired can be analyzed, communicated with other signals and imaging in real time. In general, an inductive shunting element can also be connected between the PZT patch and the NC circuit to synthesize a hybrid shunting circuit. This design can achieve the signal amplification at specific narrow band frequencies, which will be useful for frequency selection. Furthermore, the adaptive GRIN metamaterial sensing system with NC shunting circuits has several unique features that distinguish it from the charge or voltage amplification conditioning circuits used in conventional sensing systems. Due to the feedback control effect of the NC shunting circuit, the effective bending stiffness of the metamaterial can be tuned to achieve a GRIN along the waveguide, where flexural waves can be compressed and the hybrid strain amplification is visualized. However, for the conventional amplification circuits, the amplification is processed in a one-way manner, and the mechanical properties of the piezoelectric sensors are not altered. Therefore, the previous technology does not contribute to the sensor's sensitivity and SNR.

In addition, the adaptive metamaterial waveguide possesses a very high tunability and flexibility for various sensing environments and requirements by online tuning the circuit parameters. For example, the adaptive waveguide can achieve unidirectional as well as multidirectional sensing amplification by creating an asymmetric and symmetric GRIN, respectively, by properly tuning the electric control circuits. The waveguide can also be extended into 2D wave amplification in a plate structure, which will provide unprecedented wave control abilities and allow for a much wider range of applications.

It should be noted that the size of the fabricated adaptive metamaterial waveguide is very compact and light weight. Furthermore, the wave amplification can also work with discrete or even single shunted piezoelectric patches. The implementation of NC shunting circuits into piezoelectric sensors, which includes not only elastic wave sensors, but also acoustic and ultrasonic wave sensors in other fields, will have a broad impact on wave signal amplification methods.

In conclusion, enhanced flexural wave sensing through the use of an adaptive GRIN metamaterial is demonstrated, which has great potential to achieve superior performance and functionalities than the current elastic wave sensing systems. We believe that the MBSS system with shunted piezoelectric materials incorporated can open new possibilities in elastic and acoustic wave sensing applications including, but not limited to structural health monitoring, sonar imaging, acoustic systems, as well as medical instrumentation and imaging.

## Methods

**Numerical simulations of GRIN metamaterial devices.** We perform multi-physics numerical simulations considering piezoelectric and shunting circuit effects by using the commercial software COMSOL Multiphysics. In dispersion simulations, the side boundaries of the host beam along the wave propagation direction are set as symmetric, and the other two boundaries are left free to describe 1D wave propagating along the beam structure. The shunting circuit effects are prescribed in terms of impedance values in a weak form. The voltages on each of the electrodes of the PZT patches are constrained in a pointwise manner. In the frequency domain analysis, two perfectly matched layers are added and connected to the metamaterial waveguide to suppress reflected waves from boundaries. In the time domain analysis, a generalized alpha time stepping method is adopted to reach a fast solution convergence.

**Metamaterial sample preparation.** The adaptive GRIN metamaterial is fabricated by surface-bonding seven equally-spaced PZT patches (APC 851) shunted with gradient NC circuits on one side of the host beam (Aluminum 6061) to form the active waveguide (see Fig. 5a). The piezoelectric elements were positioned on the substrate beam with the help of a custom made, 3D printed mask and bonded using superglue.

**NC circuit preparation.** In the NC circuit, a 1 nF capacitor is selected for $C_0$. The resistance for $R_2$ and $R_0$ are 10 kΩ and 1 MΩ, respectively. The maximum resistance of the potentiometer $R_1$ is 100 kΩ. Seven high voltage operational amplifiers (OPA445AP) with proper protection circuits are used in all of the NC circuits. The gradient NC circuits are individually connected to the seven piezoelectric patches and powdered by a DC power supply (Instek SPD-3606). All of the electrical components are connected using prototyping breadboards.

**Experimental setup and measurements.** The low voltage tone burst signal is generated by a Tektronix AFG3022C arbitrary waveform generator and amplified by using a Krohn-Hite high voltage power amplifier and the amplified voltage signals are directly measured by a Tektronix DPO4034B digital oscilloscope connected to the output terminals of operational amplifiers in NC circuits. The out of plane velocity field on the surface of the sample is measured using a Polytec PSV-200 scanning laser vibrometer. The wave propagation properties within the GRIN metamaterial are calculated based on the velocity amplitude averaged over regions extending 300 mm on either side of the metamaterial region with a spatial resolution of approximately 1.5 mm. The velocity signal is acquired with a 128 kHz sampling frequency. For each point, we measure the response of the system to 20 consecutive tone burst signals to obtain reliable data.

## Acknowledgements

This work was supported by the Air Force Office of Scientific Research under Grant No. AF 9550-15-1-0061 with Program Manager Dr. Byung-Lip (Les) Lee.

## Author contributions

Y.Y.C. and G.L.H. developed the metamaterial enhanced sensing concept and design. Y.Y.C. conducted the analytical and numerical simulations. R.Z. and Y.Y.C. constructed the experimental setup and carried out the experiments. G.L.H., Y.Y.C. and M.V.B. wrote the manuscript. G.L.H. supervised the research and led the project.